# Boundary conditions for drift-diffusion equations in gas-discharge plasmas

V.V. Gorin[1,2], A.A. Kudryavtsev[1,3], Jingfeng Yao[1], Chengxun Yuan[1] and Zhongxiang Zhou[1]

[1]*Harbin Institute of Technology, Harbin 150001, China*
[2]*Moscow Institute of Physics and Technology, Dolgoprudny 141700, Russia*
[3]*St. Petersburg State University, St. Petersburg 198504, Russia*

This paper develops a general approach to the derivation of the boundary conditions for hydrodynamic equations for charged and neutral plasma components. It includes both a well-known classical case for pure diffusion, and considers the expressions for diffusion and drift together – for an absorbing (neutralizing) wall with partial reflection and the possible emission of plasma components. Some unclear and controversial terms found in the existing literature are clarified. Several examples of applications of the results, which illustrate the properties of boundary conditions for electrons and ions, are calculated and analyzed.

## I. INTRODUCTION

Drift-diffusion (hydrodynamic, fluid) equations are widely used [1, 2, 3, 4] to solve problems of low temperature plasma physics (here we consider situations in which the flow of the main component of plasma medium, neutral gas, is absent). These equations include

$$\frac{\partial n}{\partial t} + \operatorname{div} \mathbf{\Gamma} = s, \quad \mathbf{\Gamma} = -D\nabla n + \mathbf{V}_d n, \tag{1}$$

where $n$ is the number density of some particle species, $\mathbf{\Gamma}$ is the related flux-density vector, $s$ is the source density, $D$ is the diffusion coefficient, $\mathbf{V}_d = \mu \mathbf{E}\,\mathrm{sgn}\,q$ is the drift velocity, $q$ is the electric charge of a particle, $\mu$ is the mobility, and $\mathbf{E}$ is the electric field. These equations provide a hydrodynamic description for both charged- and neutral- particle densities in a plasma. The first equation is the particle balance equation, while the second one is an expression for the particle flux-density vector, which is the sum of the corresponding diffusion and drift flux densities.

Drift and diffusion themselves are very general in nature.

Drift is the average movement of a particle under the action of a traction force, which is balanced by the friction force. The friction force is the average energy loss per unit path length due to any dissipative events. The friction force depends on the average particle velocity. In the case of a linear dependence, a mobility factor appears. In strong fields (at low densities), when the equilibrium between traction and friction is absent, the dependence of the average velocity on the field may be nonlocal [5]. In this situation, we can formally preserve the structure of the expression for the flux density in (1), assuming that $D = 0$, and $\mathbf{V}_d$ is the average flow velocity.

Diffusion is a natural consequence of any chaotic (Brownian) particle motion [6] and usually leads the medium to equalize its concentrations.

Both of these phenomena occur both in liquids and in gases, and are not unique only to plasma particles.

A pair of equations (1) for each charged component, together with the so-called electron-energy component in the *extended fluid approach* [7], and the Poisson equation, constitute a self-consistent system of equations for plasma components in the plasma modeling problem. Eqs. (1) can be obtained for electrons and ions in a quasi-neutral plasma, in which the electric field is quite small, by averaging the Boltzmann equation. In this situation the distribution function is represented in the *Lorentz two term approximation* (LTTA) [8 - 10]; that is, the distribution of particles in the direction of motion is considered to be almost isotropic.

In large fields, the ion drift velocity often exceeds the chaotic velocity [9]. The ion velocity distribution is far from isotropic, and therefore requires a different approach. However, a drift-diffusion pair, such as in the Eq. (1), with a possible nonlinear drift or nonlocal dependence of the average velocity on the field, is also valid for large fields.

In order to obtain a unique solution when modeling a gas discharge, a boundary-value problem must be formulated, that is, *boundary conditions* (BC) should be added as restrictions on the set of solutions for each pair of equations (1). BC should be set at all boundaries of the discharge volume, including walls with weak and strong fields in their vicinity. This situation makes it necessary to find a universal formulation of the BC, which can include all kinds of fields and particles when simulating a discharge.

In the existing literature, the authors used different approaches to the formulation of BC for plasma, and the approaches differ depending on the level of understanding and complexity.

The *simplest* and roughest approach to formulation BC involves providing a zero value of the particle density at the absorbing (or neutralizing) wall, that is, a uniform Dirichlet condition [11]. If the "wall" is a plane or axis of symmetry, and the normal component of the field is zero, it can be considered as reflecting particles [12]. In this case, a uniform Neumann BC arises.

Another approach, a *primitive approach* [4], is to neglect the thickness of the distorted layer near the absorbing wall and use the "half-Maxwellian" distribution to determine the flux to the absorbing wall by integrating the flux density with the Maxwell distribution over the half of the velocity space in which the particles go to the wall. In this way, BC arise in the form $\mathbf{\Gamma} \cdot \mathbf{n} = \frac{1}{4} n \bar{v}$, $\bar{v} = \sqrt{8kT/(\pi m)}$. Since the flux density includes the terms of drift and diffusion, that is, the density and its spatial derivative, the Dirichlet condition is replaced by a more general condition of the 3rd kind, or the Robin condition [13].

It is worth noting that an approach with a *high level of understanding* assumes that the absorbing wall strongly distorts the isotropy of the distribution function near the wall, and LTTA becomes inapplicable. This follows from the formulation of the boundary condition on the absorbing wall for the particle distribution function: $f(t,\mathbf{r},\mathbf{v})=0: (\mathbf{r}\in B)\wedge(\mathbf{v}\cdot\mathbf{n}<0)$ (here $B$ is the set of boundary points, and **n** is the normal to the boundary directed outside the plasma). That is, there are no particles at the absorbing wall that move from the wall to the plasma, and the distribution function can have nonzero values only for particles moving towards the wall. But in LTTA, the distribution function is almost isotropic, and it can satisfy this condition only when the distribution function has a zero value in all directions of velocity.

However, a high-level approach (the kinetic approach) is more mathematically complicated. It is similar to the Milne problem [14], which was formulated initially for solar radiation. Sunlight propagates inside the Sun, like particle diffusion. The surface of the Sun releases light outward, so that it looks like an absorbing wall for the light propagating inside the Sun. The kinetic equation in the Milne problem was formulated for isotropic scattering of light by heavy, unmovable random centers. The solution in this study is also used for neutrons [15]. Electrons near an absorbing wall in a plasma require a different kinetic equation, since their scattering is small-angle and sharply anisotropic. The situation is different from the Milne problem, since the total cross-section of their scattering diverges. An attempt to solve this problem for electrons was made by V. V. Gorin [16].

Because of these complications, many authors [17 – 20] tended to avoid anisotropic kinetic (Boltzmann) equations and ignored the fact of distortion, trying to limit themselves to the Maxwell distribution. In this regard, an approach of *intermediate* quality arises. It can be considered to provide *effective BC*. It uses LTTA, which makes it somewhat better than a primitive approach; however, it is also quite far from logical completeness.

The idea [17] was proposed for this approach: the particle flux was divided into two types: 1) a particle flux with a positive value $v_x = \mathbf{v}\cdot\mathbf{n}$, and 2) a particle flux with a negative value $v_x$. The first type moves from plasma to the wall, and the second type – from wall to plasma. This idea can be explained by the logical difference between the two types of particles. The first type *does not depend* on the direct effect on the wall, but the second type *depends* on the reflection and emission properties of the wall.

This idea, referred to in the literature as the *two-stream approximation* [21, 22], was originally developed only for diffusion flux [17], and LTTA was used in an implicit form (by drawing pictures). This was the next step towards a more precise definition of the boundary condition compared to the primitive approach. The sum of the two terms in LTTA allowed to equate the flux from the absorbing wall to zero. This became possible due to the fact that if the sum of two terms is equal to zero, then both terms themselves may not be equal to zero, and during integration it is not necessary to remove half the space of velocities without a logical ground, since this was done in a primitive approach. This improved approach provides BC for a normal flux density of $\mathbf{\Gamma} \cdot \mathbf{n} = \frac{1}{2} n \bar{v}$.

Later, attempts were made to generalize the LTTA pure diffusion model to the drift-diffusion model. In particular, some authors [18] tried to expand the boundary condition for cases with diffusion only [17] to cases with both drift and diffusion, as well as with an emitting wall. However, instead of including drift in LTTA along with diffusion in the initial expression for the flux (because the drift flux is a component of the general flux in LTTA, the same as the diffusion flux), they added the "drift term", as they thought, to the final McDaniel expression for flux[1]. As a result, a strange quantity $a$ arose[2] in their "generalized" formulas, which turned *on* the drift flux if the drift was directed toward the wall, and turned *off* the drift flux if the drift was directed from the wall to the plasma[3]. The authors did not provide an explanation of why this non-physical switch appeared, and they simply referred to the authority of Boeuf and Pichford [4]. However, in a study by Boeuf and Pichford [4], BC were specified separately for electrons and ions, and the switch was not used.

In a study by Wilson and Shotorban [3], various boundary conditions for a fluid model were examined and compared, and it was demonstrated that the choice of BC can have a significant effect on the results of plasma modeling. Therefore, the determination of the correct BC is an urgent problem requiring further attention.

In this article, we give a generalized formula that does not require any switches. Instead, its derivation requires a consistent application of the two-stream approach, and should include both drift and diffusion terms as equitable components of the total flux, since they have equal significance for the final result. Moreover, we show *how far* one can advance in the BC formulation without using LTTA and without solving the kinetic equation near the wall.

## II. GENERAL EXPRESSIONS

Consider the general situation in which LTTA may not be applicable: the particle distribution function $f(x, \mathbf{v})$ at the boundary of the plasma $x = 0$ (the plasma is in the region $x < 0$) is unknown. However, we hope to provide plausible estimates for the quantities, that arise in this situation. Following the two-stream approach [17], we separately determine two quantities: (a) the flux density, consisting of particles moving from the plasma *to* the wall; and (b) the flux density, consisting of particles that move *from* the wall to the plasma

---

[1] See their formula (9) – against McDaniel's (10.7.1) – (10.7.3) in [17] at zero reflection factor.
[2] See [18] expression (7).
[3] See formulas (7 – 9) in their paper [18].

$$\Gamma_+ = \int_{v_x>0} d^3v f(x,\mathbf{v}) v_x, \tag{2}$$

$$\Gamma_- = \int_{v_x<0} d^3v f(x,\mathbf{v})(-v_x). \tag{3}$$

These quantities are always non-negative by definition. Therefore, for the total flux density, we have

$$\Gamma_x = \int d^3v f(x,\mathbf{v}) v_x = \int_{v_x>0} d^3v f(x,\mathbf{v}) v_x + \int_{v_x<0} d^3v f(x,\mathbf{v}) v_x = \Gamma_+ - \Gamma_-. \tag{4}$$

Particles that come from the wall into the plasma consist of both reflected particles and particles emitted by the wall:

$$\Gamma_- = R\Gamma_+ + \Gamma_{em}. \tag{5}$$

Here $R$ is the reflection coefficient[4].

In addition to Eqs. (4) and (5), we need one more independent relation between the flux densities $\Gamma_+, \Gamma_-$ in order to obtain the boundary condition. Using Eqs. (2) and (3), we obtain the expression for the sum of these flux densities:

$$\Gamma_+ + \Gamma_- = \int_{v_x>0} d^3v f(x,\mathbf{v}) v_x + \int_{v_x<0} d^3v f(x,\mathbf{v})(-v_x) = \int d^3v f(x,\mathbf{v}) |v_x| \equiv n\overline{|v_x|}. \tag{6}$$

From Eqs. (4) and (6), we can now obtain the expressions

$$\Gamma_+ = \frac{1}{2}\left(n\overline{|v_x|} + \Gamma_x\right), \tag{7}$$

$$\Gamma_- = \frac{1}{2}\left(n\overline{|v_x|} - \Gamma_x\right). \tag{8}$$

We substitute these expressions in Eq. (5)

$$\frac{1}{2}\left(n\overline{|v_x|} - \Gamma_x\right) = \frac{1}{2}\left(n\overline{|v_x|} + \Gamma_x\right)R + \Gamma_{em},$$

and then we obtain an expression for the total flux density

$$\Gamma_x = \frac{1-R}{1+R} n\overline{|v_x|} - \frac{2}{1+R}\Gamma_{em}. \tag{9}$$

Please note that the expression is independent of the applicability of LTTA. For the Maxwell distribution, in particular, we have

$$\overline{|v_x|} = \overline{v}/2, \quad \overline{v} = \sqrt{\frac{8kT}{\pi m}}, \tag{10}$$

in which division by 2 occurs due to averaging over the cosine of the angle of inclination of the velocity. The terms of drift and diffusion used in LTTA do not contribute to Eq. (10), because these terms have odd symmetry with respect to the velocity variable. But the quantity in Eq. (10) is defined

---
[4] More detail about $R$ see the Appendix.

in Eq. (6), and it has even symmetry.

In the case when ions are in a strong field, if we do not have a solution for the ion distribution function near the wall, we can estimate the quantity $\overline{|v_x|}$ (average of the absolute value of the x-component of the velocity) as the absolute value of the ion drift velocity $\overline{|v_x|} \approx |V_d|$, $V_d = \text{sgn}(q)\mu E_x$. The sign of the drift velocity $V_d$ depends on the sign of the electric charge $q$ of a particle, as well as on the direction of the electric field. For example, for electrons $V_d = -\mu E_x$, and for positive ions $V_d = \mu E_x$, μ is a mobility factor.

Given the second expression in Eqs. (1), we can define the expression for the normal component of the flux density as

$$\Gamma_x = -D\frac{\partial n}{\partial x} + nV_d. \tag{11}$$

At the wall, we can equate the expressions in Eqs. (9) and (11):

$$-D\frac{\partial n}{\partial x} + nV_d = \frac{1-R}{1+R}n\overline{|v_x|} - \frac{2}{1+R}\Gamma_{em}.$$

Here, the left side of the equation depends on the plasma parameters, and the right side depends on the properties of the wall. After transformations BC takes the form

$$D\frac{\partial n}{\partial x} + nV = \frac{2}{1+R}\Gamma_{em}. \tag{12}$$

In grouping the summands, we introduced here the quantity of velocity dimension

$$V = \frac{1-R}{1+R}\overline{|v_x|} - V_d, \tag{13}$$

which is referred to in this paper as the *Hopf velocity*. Also, for the convenience of further presentation of the work, we introduce terminology for some new quantities. BC in Eq. (12) can be represented in canonical mathematical form:

$$h\frac{\partial n}{\partial x} + n = n_{em}, \tag{14}$$

where

$$h = \frac{D}{V} \tag{15}$$

is called the *Hopf shift*, and

$$n_{em} = \frac{2}{1+R}\frac{\Gamma_{em}}{V} \tag{16}$$

is called the *Hopf emission density.*

Here $V$ can be any real number. If $V = 0$, we return back to Eq. (12) and obtain the Von Neumann condition (which is not uniform in the case of emission). The emission flux density $\Gamma_{em}$ is non-negative by definition. However, the Hopf emission density $n_{em}$, as well as the Hopf shift $h$, can take any real value, but must have the same sign, as the Hopf velocity $V$ in Eq. (13).

In the general case, the boundary condition in Eq. (14) is a non-uniform boundary condition of the 3rd kind (also called the Robin condition [13]). It has a simple geometric interpretation.

Let's draw a graph of the function $y = n(x)$. We also draw a tangent to the graph with a dashed line, which we build for the point $x = 0$, $y = n(0)$. The equation of this tangent has the form

$$y = n(0) + \frac{\partial n}{\partial x}(0) x. \qquad (17)$$

Substituting $x = h$ into Eq. (17), and taking into account the BC (14), we obtain $y = n_{em}$. That is, the tangent defined by Eq. (17) passes through the point $H(h, n_{em})$ (see Fig. 1).

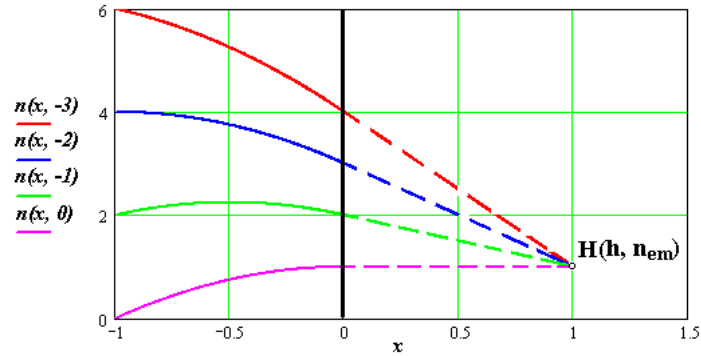

**Fig. 1.** The set of solutions $n(x, p) = 1 + p(x-1) - x^2$, $x < 0$, with a parameter $p = (\partial n / \partial x)|_{x=0} = 0, -1, -2, -3$ for the stationary one-dimensional drift-diffusion equation $\partial/\partial x(-D \partial n/\partial x + V_d n) = s$. For the image we chose the parameters $D = 1$, $V_d = 0$, $s = 2$. The solutions obey the non-uniform boundary condition of the 3rd kind (14). The tangents pass through the Hopf point $H(h, n_{em}) = (1, 1)$.

It is known [23] that the general solution of an ordinary second-order differential equation is a two-parameter set. One BC reduces the number of parameters by one, and the set of solutions becomes one-parameter. Fig. 1 illustrates this set. The normal derivative $p = (\partial n/\partial x)|_{x=0}$ of particle density at the boundary point $x = 0$ is chosen as the parameter of the set.

So, BC (14) selects a set of solutions for the drift-diffusion equation (second-order differential equation) for particles so that the tangent for each solution of the set defined on the plasma boundary passes through the shown point $H(h, n_{em})$. We call it the *Hopf point*.

The Hopf point can be in the first quadrant (for $V > 0$, as in Fig. 1), or in the third quadrant (for $V < 0$) of the plane of variables $x$, $n$, and also on the abscissa (if a uniform boundary condition is specified in the absence of emission from the wall $n_{em} = 0$; see the examples below). The choice of a unique solution to the boundary value problem from this set is carried out by a condition on another plasma boundary.

### III. COMPARING OF OUR RESULTS BY BC WITH THE RESULTS OF OTHER AUTHORS
#### A. Comparison with Boeuf and Pichford (1995) [4]

(a) Our BC for electrons, at a zero reflection coefficient $R = 0$, in the absence of emission from the wall and the Maxwell distribution of electrons, is

$$V = \frac{1}{2}\bar{v} - \mu_e E_x \approx \frac{1}{2}\bar{v}, \quad D_e = \frac{1}{3}\bar{v}\lambda, \quad h = \frac{D_e}{V} = 2\frac{D_e}{\bar{v}} = \frac{2}{3}\lambda, \tag{18}$$

$$\frac{2}{3}\lambda\frac{\partial n_e}{\partial x} + n_e = 0, \tag{19}$$

$$\Gamma_x = -D_e\frac{\partial n_e}{\partial x} - \mu_e E_x n_e = \frac{\bar{v}}{2}\left(-\frac{2}{3}\lambda\frac{\partial n_e}{\partial x} + 2\frac{V_d}{\bar{v}}n_e\right) = \frac{\bar{v}}{2}\left(n_e + 2\frac{V_d}{\bar{v}}n_e\right) \approx \frac{1}{2}n_e\bar{v}. \tag{20}$$

In the above equations, $\lambda$ is the mean free path of an electron. Here we neglect the electron drift velocity $V_d = -\mu_e E_x$ compared to its thermal velocity $\bar{v} = \sqrt{8kT_e/(\pi m_e)}$.

In the study by Boeuf and Pichford [4][5], the flux density is $\Gamma_x = \frac{1}{4}n_e\bar{v}$.

Thus, our electron flux density at the boundary is two times higher than in the study of Boeuf and Pichford [4]. This discrepancy is a consequence of the older one-term primitive approach that was used in this study, compared to the two-stream approach that is used in the later literature and here in our study.

(b) Our BC for positive ions in a strong field, with a zero reflection coefficient $R = 0$, the absence of emission from the wall and neglecting the number of ions moving against the electric force, has the form

$$V = \overline{|v_x|} - V_d \approx |V_d| - V_d = \begin{cases} 0, & E_x > 0; \\ 2|V_d|, & E_x < 0. \end{cases} \tag{21}$$

At $E_x > 0$, our result is

$$\frac{\partial n_i}{\partial x} = 0, \quad \Gamma_x = n_i V_d. \tag{22}$$

When the electric field has a positive $x$-component $E_x > 0$, , and the ion drift is directed from plasma to the wall, the Hopf velocity is estimated to be very low. Therefore, $h = D/V \to \infty$, and we have the Neumann BC for ion density. The ion flux determined by Eq. (11) is estimated as a drift flux.

---
[5] Page 1379 Eq. (15).

And conversely, if $E_x < 0$, our result is

$$h = \frac{D_i}{V} = \frac{D_i}{2|V_d|} \approx 0. \quad n_i \approx 0, \quad \Gamma_x = n_i V_d \approx 0. \tag{23}$$

Thus, when the electric field has a negative *x*-component, and the ion drift is directed from the wall to plasma, the Hopf velocity is estimated to be of large value ($V \approx 2|V_d|$) in comparison with the thermal velocity of the ions. Therefore, the Hopf shift $h = D/V \approx D/(2|V_d|) \approx 0$ is much smaller than the mean free path of ions $\lambda_i$, and we have the Dirichlet BC for the ion density. The ion flux determined by Eq. (11) is estimated as merely a diffusion flux, i.e., it is negligible.

Therefore, we conclude that the Boeuf and Pichford result [4][6] for ions coincides with ours. That is, no switch, as in the study of Hagelaar and others [18][7], is required to have this BC.

B. Comparison with Hagelaar and others (2000) [18]

Our BC result for the total flux under the assumption of the Maxwell distribution and in the absence of emission is

$$\Gamma_x = \frac{1-R}{1+R} \cdot \frac{1}{2} n \bar{v}. \tag{24}$$

Hagelaar and others [18][8] give the formula (in our designations):

$$\Gamma_x = \frac{1-R}{1+R}\left[(2a-1)\mathrm{sgn}(q)\mu E_x n + \frac{1}{2}n\bar{v}\right] = \frac{1-R}{1+R}\left[(2a-1)nV_d + \frac{1}{2}n\bar{v}\right].$$

Their expression for flux density has an additional *unnecessary* term with a switch $a = 0; 1$, the value of which depends on the direction of drift. It can be eliminated in all their formulas by substitution $a = \frac{1}{2} = \mathrm{const}$. But the possibility of such a correction is more an accident than a logical conclusion.

Similar situation is with electron emission. Our result is

$$\Gamma_x = \frac{1-R_e}{1+R_e} \cdot \frac{1}{2} n_e \bar{v}_e - \frac{2}{1+R_e} \Gamma_{em}, \tag{25}$$

but in their result[9], the non-physical switch *a* had arisen:

$$\Gamma_x = \frac{1-R_e}{1+R_e}\left[-(2a_e - 1)\mu_e E_x n_e + \frac{1}{2}n_e \bar{v}_e\right] - \frac{2}{1+R_e}\Gamma_{em}.$$

---

[6] See page 1379 Eq. (18) and the text after formula.
[7] See Eq. (7).
[8] See page 1453 Eq. (11).
[9] See [18] page 1453 Eq. (13).

The unnecessary term with the switch *a* had arisen here on the same reason: the drift term was not included in the original kinetic expression along with the diffusion term, and the reasonable derivation of the formula for the flux density was replaced by some fantasy.

## IV. EXAMPLES

Let's now give specific examples. For simplicity of calculations and clarity of results, we restrict ourselves to considering one-dimensional stationary problems of drift and diffusion along the *x*-coordinate. Along the other Cartesian coordinates *y* and *z*, we assume that the solution is uniform (that is, it does not depend on these coordinates).

### A. Neutral atoms in the ground state in a gas discharge

The source of atoms is the negatively charged wall, where neutralization of positive ions occurs. A drain of atoms is ionization in a plasma volume. Equations in plasma volume are

$$\text{div}\,\mathbf{\Gamma} = -s_{ion}, \quad \mathbf{\Gamma} = -D\nabla n. \tag{1.1}$$

The one-dimensional variant is

$$s_{ion} = 1, \quad D = 1: \quad \frac{d\Gamma_x}{dx} = -1, \quad \Gamma_x = -\frac{dn}{dx}; \quad -1 < x < 0. \tag{1.2}$$

The general solution is

$$\Gamma_x(x) = \Gamma_x(0) - x, \quad n(x) = n(-1) + \frac{1}{2}(x^2 - 1) - \Gamma_x(0)(x+1). \tag{1.3}$$

The parameters in Eqs. (12 – 15) are

$$R = 1, \quad \Gamma_{em} = \int_{-1}^{0} dx\, s_{ion} = 1, \quad V = 0. \tag{1.4}$$

The boundary condition in Eq. (12) gives

$$\frac{dn}{dx}(0) = \Gamma_{em} = 1. \tag{1.5}$$

The set of solutions obeying BC (12) has the form

$$n(x) = n(-1) + \frac{1}{2}(x+1)^2, \tag{1.6}$$

where the value of *n* at *x* = -1 is a parameter of the set.

In this example, the Hopf point $H(h, n_{em})$ goes to infinity in a direction in which the ratio $n_{em}/h = \Gamma_{em}/D$ remains constant (see Fig. 2).

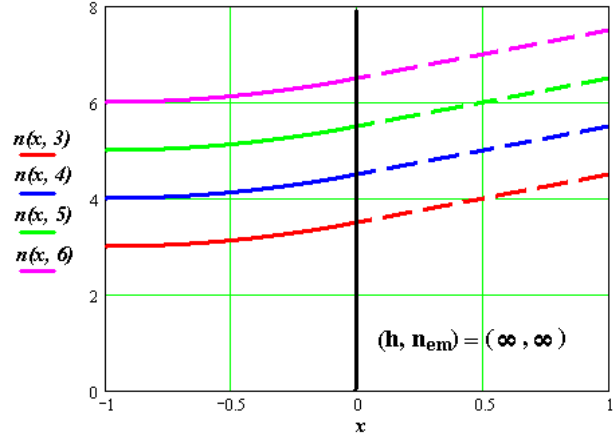

**Fig. 2.** Solutions for the density of neutral atoms have the same slope of the tangent at the right boundary (at the cathode). The entry of atoms from the cathode surface into the plasma is compensated by their disappearance during ionization in the depth of the plasma volume.

### B. Positive ions in a dark DC discharge caused by external radiation

To simplify the solution, we consider the space charge small enough to neglect it, and we assume that the electric field is uniform. Let the ionization source be spatially uniform, as from external radiation. Then we have the equations

$$\operatorname{div}\mathbf{\Gamma} = s_{ion}, \quad \mathbf{\Gamma} = -D\nabla n + \mu \mathbf{E} n. \tag{2.1}$$

One-dimensional statement of the problem is

$$s_{ion} = 1, \quad \mu E_x = 1: \quad \frac{d\Gamma_x}{dx} = 1, \quad \Gamma_x = -D\frac{dn}{dx} + n; \quad 0 < x < 1. \tag{2.2}$$

The anode is located on the left at $x = 0$, and the cathode is on the right at $x = 1$. As a parameter, we leave the diffusion coefficient, which should be small when we consider a strong electric field. A general solution is

$$\Gamma_x(x) = \Gamma_x(0) + x, \tag{2.3}$$

$$\frac{dn}{dx} - \frac{1}{D}n = -\frac{1}{D}(\Gamma_x(0) + x), \quad 0 < x < 1;$$

$$n(x) = n(0)\exp\left(\frac{x}{D}\right) - \frac{1}{D}\int_0^x dx'(\Gamma_x(0) + x')\exp\left(\frac{x-x'}{D}\right),$$

$$n(x) = \exp\left(\frac{x}{D}\right)\left(n(0) + \int_0^x d\left((\Gamma_x(0) + x' + D)\exp\left(\frac{-x'}{D}\right)\right)\right),$$

and finally we obtain

$$n(x) = A\exp\left(\frac{x}{D}\right) + B\left[1 - \exp\left(\frac{x}{D}\right)\right] + x, \tag{2.4}$$

$$A = n(0), \quad B = \Gamma_x(0) + D.$$

Emission is absent both from the anode and from the cathode, therefore $\Gamma_{em} = 0$. We neglect the reflection of ions from the cathode $R_c = 0$. If we assume a large drift of ions in comparison with diffusion of ions [10] $D/|V_d| \ll 1$, then an approximate estimate is as follows $\overline{|v_x|} \approx |V_d| = \mu E_x = 1$. Then, for the Hopf velocity (13) in the anode we obtain

$$V_{(a)} \approx \frac{1-R}{1+R}V_d + V_d = \frac{2}{1+R}V_d, \tag{2.5}$$

and in the cathode we obtain

$$V_{(c)} \approx \frac{1-R_c}{1+R_c}V_d - V_d \approx 0. \tag{2.6}$$

That is, in accordance with Eq. (15), $h_{(a)} \approx 0$ at the anode, and $h_{(c)} \approx \infty$ at the cathode. Thus, we obtain the Dirichlet boundary condition $n \approx 0$ at the anode, and the Neumann boundary condition $dn/dx \approx 0$ at the cathode. These conditions are very approximate, especially at the cathode, because our estimates are very rough in the absence of an ion distribution function at the boundary. According to our estimates, the solution to the Eqs. (2.4) takes the form

$$n(x) = x + (n(1) - 1)\frac{1 - \exp(x/D)}{1 - \exp(1/D)}. \tag{2.7}$$

You can see the behavior of the solution in Fig. 3.

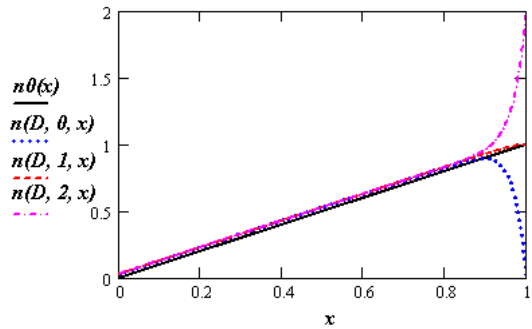

**Fig. 3.** The behavior of the solution at $D = 0.03$, and for different values of the parameter $n(1)$ at the cathode (on the right side). The solid line is the unperturbed solution, at $D = 0$. The dashed line corresponds to the solution with the Neumann condition $dn/dx\,(x=1) = 0$ at the cathode. The dashed-dotted and dotted curves are possible solutions due to the poor definiteness of the BC at the cathode.

The small diffusion coefficient of ions $D$ at the highest derivative in the drift-diffusion equation leads to a *singularly perturbed problem* [24]: if for $D = 0$ a problem is considered unperturbed, then its formulation contains fewer boundary conditions; in our example, there is one condition in the anode. The transition to a small value of $D > 0$ should have a weak effect on the solution, on the one hand; on the other hand, it increases the number of boundary conditions necessary for the solution to be unique. The graphs in Fig. 3 show how these conflicting qualities can be combined. Namely, a small value of $D$ has practically no effect in the vicinity of the anode and in the depth of the plasma volume; a special region of influence of a small value of $D$ and the second BC is in the vicinity of the cathode, where all the variety of solutions that is provided by this condition is concentrated. The Hopf shift at the cathode is actually *indefinite*, since it is equal to the ratio of small quantities: the diffusion coefficient $D$ and the Hopf velocity $V$. The accuracy of determining this ratio is *worse* than the accuracy $D$ due to the lack of an ion distribution function. Here, the Neumann BC has a formal rather than a real meaning, because there simply must be something in terms of a correct mathematical formulation.

---

[10] The width of the discharge gap between the anode and cathode here is taken as a unit of length.

### C. Electrons in a dark self-sustained discharge

We use this discharge to simplify the consideration, neglecting the space charge, and we assume that the electric field is uniform. The density of the ionization source in the local ionization model is approximately proportional to the absolute value of the electron flux density:

$$\text{div}\,\mathbf{\Gamma} = s_{ion} = \alpha|\mathbf{\Gamma}|, \quad \mathbf{\Gamma} = -D\nabla n - \mu\mathbf{E}n. \tag{3.1}$$

One-dimensional statement of the problem is

$$\frac{d\Gamma_x}{dx} = \alpha|\Gamma_x|, \quad \Gamma_x = -D\frac{dn}{dx} - \mu E_x n; \quad 0 < x < d. \tag{3.2}$$

Let the anode be on the left at $x = 0$, and the cathode be on the right at $x = d$. The electron flux is then negative:

$$\frac{d\Gamma_x}{dx} = -\alpha\Gamma_x. \tag{3.3}$$

Substituting the flux into the balance equation results in

$$D\frac{d^2n}{dx^2} + (V_d + \alpha D)\frac{dn}{dx} + \alpha V_d n = 0; \quad 0 < x < d. \quad V_d = \mu E_x. \tag{3.4}$$

Based on our assumptions, this is an ordinary differential equation with constant coefficients. Its general solution is a linear combination of exponentials:

$$n(x) = Ae^{\lambda_1 x} + Be^{\lambda_2 x}, \tag{3.5}$$

the increments in which are solutions of the algebraic equation:

$$D\lambda^2 + (V_d + \alpha D)\lambda + \alpha V_d = 0,$$

$$\lambda_{1,2} = \frac{-(V_d + \alpha D) \pm \sqrt{(V_d + \alpha D)^2 - 4\alpha D V_d}}{2D}, \quad (3.6)$$

$$\lambda_{1,2} = \frac{-(V_d + \alpha D) \pm |V_d - \alpha D|}{2D} = -\frac{V_d}{D}, -\alpha.$$

Therefore, the general solution has the form

$$n(x) = A\exp\left(-\frac{V_d}{D}x\right) + B\exp(-\alpha x). \quad (3.7)$$

At the cathode is an emission source $\Gamma_{em} = \gamma \Gamma_i > 0$. Suppose that the electron velocity distribution is close to isotropic and Maxwellian. Then we have

$$\overline{|v_x|} = \frac{1}{2}\overline{v}, \quad \overline{v} = \sqrt{\frac{8kT_e}{\pi m_e}}. \quad (3.8)$$

We select the primary (input) parameters: gas is Argon,

$$R = 0, \quad \frac{kT_e}{e} = 1\,\text{eV}, \quad d = 4\,\text{cm}, \quad U = 250\,\text{V}, \quad P = 1\,\text{Torr}, \quad T = 300°\text{K}, \quad n_{em} = 10^{10}\,\text{m}^{-3}. \quad (3.9)$$

The secondary (derived) parameters are

$$N_{Ar} = \frac{P}{kT} = 3.219 \cdot 10^{22}\,\text{m}^{-3}, \quad E = \frac{U}{d} = 6250\,\text{V/m}, \quad \overline{v} = \sqrt{\frac{8kT_e}{\pi m_e}} = 6.692 \cdot 10^5\,\text{m/s},$$

$$\mu_e = 33\,\text{m}^2/(\text{s}\cdot\text{V}), \quad D = \mu_e \frac{kT_e}{e} = 33\,\text{m}^2/\text{s}, \quad V_d = \mu_e E = 2.063 \cdot 10^5\,\text{m/s}, \quad \lambda_{free} = \frac{3D}{\overline{v}} = 0.148\,\text{mm},$$

$$\alpha = 100.955\,\text{m}^{-1}, \quad \alpha d = 4.038, \quad d/\lambda_{free} = 270.396, \quad \gamma^{-1} = e^{\alpha d} - 1 = 55.724.$$

$$(3.10)$$

Criterion for small space charge is

$$\Delta E = E_c - E_a \sim \frac{e}{\varepsilon_0} n_{em} e^{\alpha d} \cdot d = 411.071\,\text{V/m}, \quad \Delta E/E = 0.066 \ll 1. \quad (3.11)$$

The Hopf parameters at the cathode are

$$V_{(c)} = \frac{1-R}{1+R}\overline{|v_x|} - (-V_d) = \frac{\overline{v}}{2} + V_d = 5.409 \cdot 10^5\,\text{m/s}, \quad h_{(c)} = \frac{D}{V_{(c)}} = 0.061\,\text{mm}, \quad \frac{h_{(c)}}{\lambda_{free}} = 0.412. \quad (3.12)$$

The Hopf parameters at the anode are

$$V_{(a)} = \frac{1-R}{1+R}\overline{|v_x|} - V_d = \frac{\overline{v}}{2} - V_d = 1.284 \cdot 10^5\,\text{m/s}, \quad h_{(a)} = \frac{D}{V_{(a)}} = 0.2571\,\text{mm}, \quad \frac{h_{(a)}}{\lambda_{free}} = 1.738. \quad (3.13)$$

The boundary condition from Eq. (14) at the cathode is

$$h_{(c)}\frac{dn}{dx} + n = n_{em}. \quad (3.14)$$

The boundary condition from Eq. (14) at the anode is

$$-h_{(a)} \frac{dn}{dx} + n = 0. \tag{3.15}$$

We substitute these conditions into the general solution. To do this, let's first calculate the density gradient:

$$\frac{dn}{dx}(x) = -A \frac{V_d}{D} \exp\left(-\frac{V_d}{D} x\right) - B\alpha \exp(-\alpha x). \tag{3,16}$$

Substituting the boundary conditions, we obtain a system of two linear algebraic equations for calculating the constants $A$ and $B$:

$$\begin{cases} A\left(\exp\left(-\frac{V_d}{D}d\right) - h_{(c)} \frac{V_d}{D} \exp\left(-\frac{V_d}{D}d\right)\right) + B\exp(-\alpha d)(1-\alpha h_{(c)}) = n_{em}, \\ A\left(1 + h_{(a)} \frac{V_d}{D}\right) + B(1 + \alpha h_{(a)}) = 0. \end{cases} \tag{3.17}$$

We find a solution using the Cramer rule [25]. Determinants of the system are equal to

$$\Delta = \det \begin{bmatrix} \left(\exp\left(-\frac{V_d}{D}d\right) - h_{(c)} \frac{V_d}{D} \exp\left(-\frac{V_d}{D}d\right)\right) & \exp(-\alpha d)(1-\alpha h_{(c)}) \\ \left(1 + h_{(a)} \frac{V_d}{D}\right) & (1 + \alpha h_{(a)}) \end{bmatrix} = -0.046, \tag{3.18}$$

$$\Delta_A = \det \begin{bmatrix} n_{em} & \exp(-\alpha d)(1-\alpha h_{(c)}) \\ 0 & (1 + \alpha h_{(a)}) \end{bmatrix} = 1.026 \cdot 10^{10} \text{ m}^{-3}, \tag{3.19}$$

$$\Delta_B = \det \begin{bmatrix} \left(\exp\left(-\frac{V_d}{D}d\right) - h_{(c)} \frac{V_d}{D} \exp\left(-\frac{V_d}{D}d\right)\right) & n_{em} \\ \left(1 + h_{(a)} \frac{V_d}{D}\right) & 0 \end{bmatrix} = -2.607 \cdot 10^{10} \text{ m}^{-3}. \tag{3.20}$$

According to the Cramer rule, we get

$$A = \frac{\Delta_A}{\Delta} = -2.246 \cdot 10^{11} \text{ m}^{-3}, \quad B = \frac{\Delta_B}{\Delta} = 5.708 \cdot 10^{11} \text{ m}^{-3}. \tag{3.21}$$

The calculation results of the above formulas obtained using the Mathcad package are presented in Fig. 4 – 7.

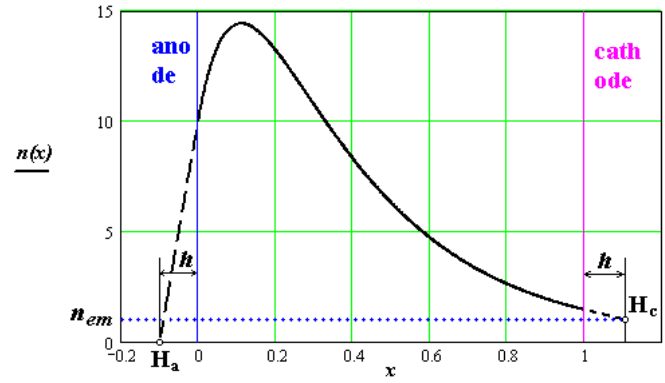

**Fig. 4.** A sketch of the behavior of the solution for electron density. Anode is at $x = 0$, and the cathode is at $x = 1$. Parameter values are $\alpha = 3$, $V_d/D = 10$, $h = 0.1$, $n_{em} = 1$.

The real physical system, as usual, has rather stiff parameters, which makes it difficult to provide an illustrated interpretation. Fig. 4 presents a sketch of our idea of how the Hopf points are located at the anode and cathode. Fig. 5 provides an illustration of the behavior of electron density at real values of physical parameters for Argon gas in a planar capacitor (see Eqs. (3.9) and (3.10)). Our estimates (3.11) show that the space charge at the chosen parameters is small enough to neglect it and exclude its effect on the electric field inside the capacitor. Under this condition, our analytical solution (3.7) and the calculations of its coefficients (3.17 – 3.21) are valid. We consider this calculation to be a mathematical simulation of dark glow.

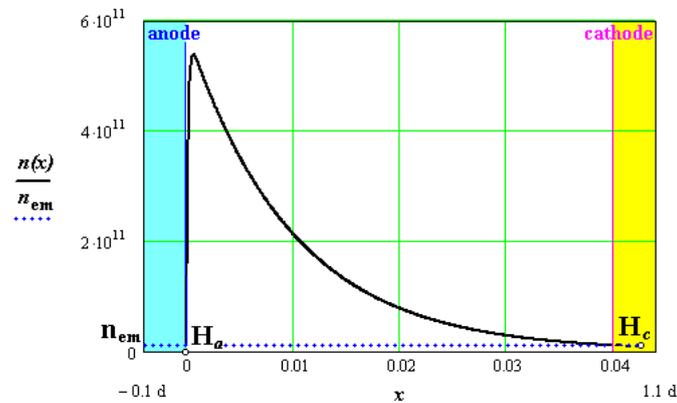

**Fig. 5.** The calculated behavior of the electron density in Argon – common picture.

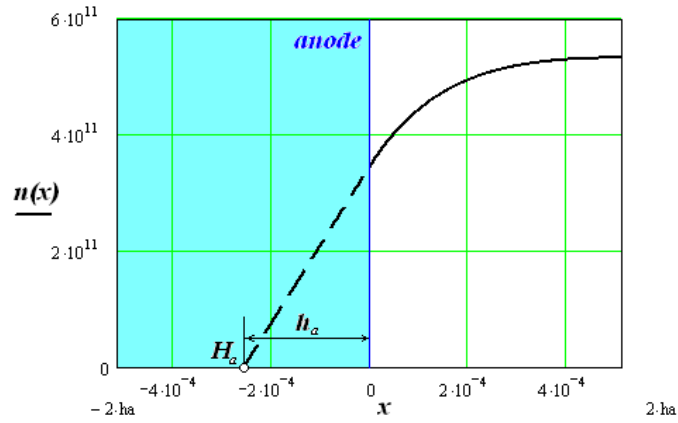

**Fig. 6.** The calculated behavior of the electron density in Argon in the vicinity of the anode.

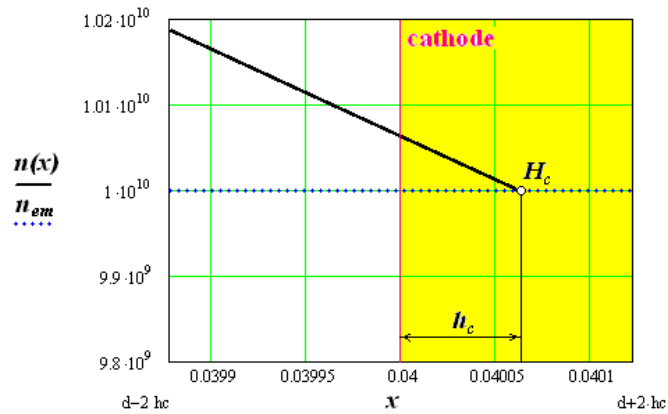

**Fig. 7.** The calculated behavior of the electron density in Argon in the vicinity of the cathode.

Fig. 5 shows that the electron density increases exponentially from the cathode to the anode (from right to left) in the Townsend avalanche. Near the anode, the density decreases due to diffusion of electrons onto the anode. The Hopf points $H_a$ and $H_c$ are shown behind the discharge walls. They determine the Robin BC at the anode and cathode.

Behind the anode (see the enlarged image in Fig. 6), the Hopf point $H_a\left(-h_{(a)}, 0\right)$ is located on the abscissa axis, since there is no electron emission from the anode. A uniform Robin BC is present here.

Behind the cathode (see the enlarged image in Fig. 7), the Hopf point $H_c\left(d + h_{(c)}, n_{em}\right)$ rises above the abscissa axis, since there is a source of electron emission at the cathode. A non-uniform Robin BC is present here.

As can be seen from the common picture in Fig. 5, the position of the Hopf point on the anode does not noticeably differ from the boundary of the anode; therefore, the Dirichlet BC $n(0) = 0$ in this case does not seem to be a significant error.

## D. Atmosphere of the Earth

This example is not about a discharge in a plasma, but it gives us the case of a negative value of the Hopf shift.

For simplicity, we assume that the surface of the Earth fully reflects the molecules of the atmosphere ($R = 1$). In addition, there is no adsorption or resorption of gases. The vertical flux is therefore zero, and we have the equation

$$\Gamma_x = -D\frac{dn}{dx} + \mu g n = 0. \tag{4.1}$$

Here, the acceleration of gravity $g$ is similar to the electric field $E$, and the gravitational mass of the molecule $m$ is similar to the electric charge. A solution of the equation is

$$n(x) = n(0)\exp\left(\frac{\mu g}{D}x\right). \tag{4.2}$$

The Einstein relation between mobility and diffusion gives

$$\frac{D}{\mu} = \frac{kT}{m}. \tag{4.3}$$

Substituting this into Eq. (4.2) we obtain the well-known barometric formula [26]

$$n(x) = n(0)\exp\left(\frac{mgx}{kT}\right). \tag{4.4}$$

The Hopf quantities are

$$V = \frac{1-R}{1+R}\overline{|v_x|} - V_d = -V_d = -\mu g,$$
$$h = \frac{D}{V} = -\frac{D}{V_d} = -\frac{D}{\mu g} = -\frac{kT}{mg}. \tag{4.5}$$

Thus, the barometric formula can be rewritten in the form (see Fig. 8)

$$n(x) = n(0)\exp\left(-\frac{x}{h}\right). \tag{4.6}$$

Here we use an analogy that compares the behavior of a neutral molecule in Earth's gravity with a charged particle in an electric field. The gravitational mass of a molecule is similar to an electric charge, and the acceleration of gravity is similar to an electric field. In this analogy, mobility and diffusion coefficients arise, and the Einstein relation also has the gravitational mass of molecule $m$ instead of the electronic charge $e$.

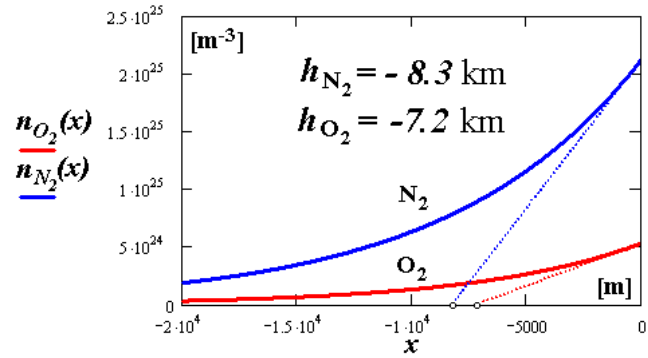

**Fig. 8.** The distribution of oxygen and nitrogen molecule densities in the Earth's atmosphere over the height -$x$ at absolute temperature $T = 273.15$ K. The Hopf points are indicated by small circles on the abscissa axis at $x = -8$ km for nitrogen molecules, and at $x = -7$ km – for oxygen molecules.

The distribution of the density of molecules here is the result of the equilibrium of the process of the drift of molecules down to the Earth's surface (in Fig. 8 it is from left to right), and the process of their diffusion upward (in Fig. 8 it is from right to left). The Hopf velocity in Eq. (13) $V = -V_d$ has a negative value because molecules reflect on the Earth's surface ($R = 1$), and the first term in Eq. (13) vanishes.

### E. Quasineutral plasma near the wall at low pressure

In this example, a quasineutral plasma is bounded by a cold wall through some charged sheath. It is well known that the diffusion and mobility coefficients of electrons in a plasma are much larger than that of ions, therefore, when a plasma begins to contact a cold electrically insulated wall, electrons are absorbed by the wall faster than positive ions, and the wall acquires a negative electric charge. The concentration of positive ions near the wall becomes greater than the concentration of electrons, and near the wall a positive electric volumetric charge arises, which forms a charged sheath between the plasma and the wall.

First, we estimate the thickness of this sheath in comparison with the mean free paths for ions and electrons.

We take some typical primary parameters for a glow discharge in Argon: gas pressure $P = 0.01$ Torr; gas temperature $T = 300$ K; ion temperature $T_i = 300$ K; electron temperature $kT_e = 1$ eV; relative plasma concentration $\eta = 10^{-6}$; elastic ion scattering cross-section $\sigma_i \approx 10^{-18}$ m$^2$; elastic electron scattering cross-section $\sigma_e \approx 10^{-20}$ m$^2$.

Secondary Parameters are as follows:
concentration of neutral atoms $N = P/kT = 3.2 \cdot 10^{20}$ m$^{-3}$; plasma concentration (electrons or ions)

$n_e = n_i \equiv n = N\eta = 3.2 \cdot 10^{14} \, \text{m}^{-3}$; electron mean free path $l_e = 1/(N\sigma_e) = 0.3 \, \text{m}$; ion mean free path $l_i = 1/(N\sigma_i) = 3 \cdot 10^{-3} \, \text{m}$; the Debye length $D = \left( \dfrac{e^2 n}{\varepsilon_0 kT_i} + \dfrac{e^2 n}{\varepsilon_0 kT_e} \right)^{-\frac{1}{2}} = 6.574 \cdot 10^{-5} \, \text{m}$; the electron Debye length $D_e = \sqrt{\dfrac{\varepsilon_0 kT_e}{e^2 n}} = 4.141 \cdot 10^{-4} \, \text{m}$.

As we can see, the Debye length of an electron, which estimates the thickness of a charged sheath, is less than the mean free path of an electron and an ion.

In the situation of a small sheath thickness compared with the mean free path, a hydrodynamic model of a charged sheath was developed (see Riemann study [27]). In this model, the *electron distribution function* (EDF) in the sheath is estimated as the Maxwell-Boltzmann distribution (the EDF obeys the stationary Vlasov kinetic equation and transforms into a Maxwell distribution in the plasma region). The behavior of ions is described as the stationary hydrodynamic flow of a cold ionic liquid using the continuity equation and the Euler equation under the action of electric force and without friction (viscosity). The Poisson equation for the electrostatic field closes the system of ordinary differential equations in the sheath:

$$n_e(x) = n_e(0) \exp\left( \dfrac{e\varphi(x)}{kT_e} \right), \tag{5.1}$$

$$n_i v_i = \Gamma, \tag{5.2}$$

$$v_i \dfrac{dv_i}{dx} = \dfrac{e}{m_i} E, \tag{5.3}$$

$$\dfrac{dE}{dx} = \dfrac{e}{\varepsilon_0}(n_i - n_e), \tag{5.4}$$

$$E = -\dfrac{d\varphi}{dx}. \tag{5.5}$$

For the Cauchy problem for this system of equations we choose the initial point of integration at the "plasma-sheath" boundary[11]. At this point, we set the electric potential to zero:

$$\varphi(0) = 0. \tag{5.6}$$

The $x$-axis we direct to the wall.

The solution of the Eqs. (5.2), (5.3) is

$$v_i(x) = \sqrt{v_i^2(0) - \dfrac{2e}{m_i}\varphi(x)}, \tag{5.7}$$

---

[11] In the literature [27], total sheath is divided on two areas: "presheath" and "sheath". We do not do this.

$$n_i(x) = \frac{\Gamma}{\sqrt{v_i^2(0) - \frac{2e}{m_i}\varphi(x)}}. \tag{5.8}$$

Substitution of the Eqs. (5.1) and (5.8) into Eq. (5.4) gives the nonlinear 2$^{nd}$ order equation:

$$-\frac{d^2\varphi}{dx^2} = \frac{e}{\varepsilon_0}\left(\frac{\Gamma}{\sqrt{v_i^2(0) - \frac{2e}{m_i}\varphi}} - n_e(0)\exp\left(\frac{e\varphi}{kT_e}\right)\right). \tag{5.9}$$

Four constants: $\Gamma, n_e(0), v_i(0), E(0)$ determine the solution to the Cauchy problem for Eq. (5.9). Constants obey conditions:

1) Quasineutrality of plasma --

$$\begin{cases} n_i(0) = n_e(0) \equiv n, \\ \frac{dn_i}{dx}(0) = \frac{dn_e}{dx}(0) \equiv n'_x. \end{cases}$$

Substitution of Eqs. (5.1) and (5.8) gives the expressions:

$$v_i(0) = \sqrt{\frac{kT_e}{m_i}}. \tag{5.10}$$

$$\Gamma = n\sqrt{\frac{kT_e}{m_i}}, \tag{5.11}$$

2) Ambipolar diffusion in the plasma --

$$\Gamma = -D_a n'_x, \tag{5.12}$$

$$D_a = \frac{1}{e}\left(\frac{1}{\mu_e} + \frac{1}{\mu_i}\right)^{-1}(kT_e + kT_i) \approx \frac{kT_e}{e}\mu_i; \tag{5.13}$$

$$E(0)n = \left(\frac{\mu_i}{D_i} + \frac{\mu_e}{D_e}\right)^{-1}\left(\frac{1}{D_i} - \frac{1}{D_e}\right)\Gamma \approx \frac{\Gamma}{\mu_i}. \tag{5.14}$$

Using Eqs. (5.11) and (5.14) we obtain:

$$-\frac{d\varphi}{dx}(0) \approx \frac{1}{\mu_i}\sqrt{\frac{kT_e}{m_i}}. \tag{5.15}$$

This is the second initial condition (together with the first in Eq. (5.6)) for the Cauchy problem for the Eq. (5.9).

Combining Eqs. (5.11) and (5.12) we obtain:

$$h^* n'_x + n = 0, \tag{5.16}$$

$$h^* = D_a\sqrt{\frac{m_i}{kT_e}}. \tag{5.17}$$

Thus, we obtain a Hopf shift, however, not relative to the wall, but relative to the "plasma-sheath" boundary. The true Hopf shift is

$$h = h^* - x_w, \qquad (5.18)$$

where $x_w$ is the sheath thickness or the coordinate of the wall surface.

As we can see from the Eqs. (5.17), (12) and (14), the quantity (5.10) could be the Hopf velocity for this example, if the boundary "plasma-sheath" were a true wall. But since the thickness of the sheath is a positive value, by the formula (5.18) we have $h < h^*$, and from Eq. (12) we have $V > \sqrt{kT_e/m_i}$. This means that the true Hopf velocity $V$ satisfies the Bohm criterion[12].

The wall coordinate $x_w$ can be estimated by equating the ion flux Eq. (5.11) to the electron flux, and the electron flux can be estimated as

$$\Gamma_e(x) = \frac{1}{4} n_e(x) \bar{v}_e = \frac{1}{4} n_e(x) \sqrt{\frac{8kT_e}{\pi m_e}}. \qquad (5.19)$$

This expression differs from the usual McDaniel estimate [17] in LTTA (with a coefficient of 1/2) for the flux of particles onto an absorbing wall that have a Maxwell distribution (see our Section III Eq. (20)) in a situation where the mean free path of a particle is much less than the inhomogeneity of the space charge, and the drop in the electric potential between two subsequent collisions is lower than the electron temperature. In our case, the *free collisionless* motion of electrons[13] through the sheath holds the Maxwell distribution for electrons going from the plasma to the wall, and the zero distribution for electrons in the opposite direction. The bulk of electrons is reflected in the sheath by an electric field and returns back to the plasma due to a lack of energy. Thus, here we have a "half-Maxwellian" distribution for electrons near the wall as the best estimate.

Equating to the flux Eq. (5.11) we obtain:

$$\frac{e\varphi(x_w)}{kT_e} = \ln\sqrt{2\pi \frac{m_e}{m_i}}. \qquad (5.20)$$

For further convenience, let's convert our quantities into dimensionless form. We choose the quantity defined by Eq. (5.16) as a unit of length and introduce the dimensionless coordinate

$$\xi = x/h^*. \qquad (5.21)$$

---

[12] See the Riemann study [27] Eq. (11).

[13] The general solution of the Vlasov equation $v_x \frac{\partial f_e}{\partial x} - \frac{e}{m_e} E(x) \frac{\partial f_e}{\partial v_x} = 0$ is an arbitrary function of the total energy $\varepsilon = \frac{m_e}{2} v_x^2 - e\varphi(x)$. We can also choose the solution $f_e(x, v_x) = C\left(1 - \theta(-v_x)\theta(\varepsilon - \varepsilon_w)\right)\exp(-\varepsilon/kT_e)$ that satisfies the condition of the absorbing wall. This is the Maxwell-Boltzmann distribution truncated to zero for phase trajectories going from the wall to the plasma. This gives a "half-Maxwellian" distribution on the wall at $x = x_w$, $\varepsilon_w \equiv -e\varphi(x_w)$.

We also introduce the following dimensionless quantities:

$$\phi(\xi) = \frac{e\varphi(\xi h^*)}{kT_e}, \quad \bar{n}_e(\xi) = n_e(\xi h^*)/n, \quad \bar{n}_i(\xi) = n_i(\xi h^*)/n, \quad \bar{v}_i(\xi) = v_i(\xi h^*)\sqrt{\frac{m_i}{kT_e}}. \quad (5.22)$$

Being constructed from these quantities, the Poisson equation (5.9) is rewritten in the form

$$-\frac{d^2\phi}{d\xi^2} = p\left(\frac{1}{\sqrt{1-2\phi}} - e^\phi\right). \quad (5.23)$$

Here the dimensionless parameter $p$ is equal to

$$p = \frac{e^2 m_i n}{\varepsilon_0 (kT_e)^2} D_a^2 \approx \frac{m_i}{\varepsilon_0} \mu_i^2 n. \quad (5.24)$$

The boundary condition Eq. (5.15) takes the form:

$$\frac{d\phi}{d\xi}(0) = -q. \quad (5.25)$$

$$q = \frac{eD_a}{\mu_i kT_e} = \left(1 + \frac{\mu_i}{\mu_e}\right)^{-1}\left(1 + \frac{T_i}{T_e}\right) \approx 1. \quad (5.26)$$

Another boundary condition Eq. (5.20), which determines the position of the wall, gives the equation

$$\phi(\xi_w) = \ln\sqrt{2\pi\frac{m_e}{m_i}} \equiv \phi_w. \quad (5.27)$$

The Cauchy problem for equation (5.23) with conditions $\phi(0) = 0$ and Eq. (5.25) can be solved analytically:

$$\left(\frac{d\phi}{d\xi}\right)^2 - 2p\left(\sqrt{1-2\phi} + e^\phi\right) = \text{const} = q^2 - 4p,$$

$$\frac{d\phi}{d\xi} = -\sqrt{q^2 + 2p\left(\sqrt{1-2\phi} + e^\phi - 2\right)},$$

$$\xi = -\int_0^\phi \frac{d\phi'}{\sqrt{q^2 + 2p\left(\sqrt{1-2\phi'} + e^{\phi'} - 2\right)}}. \quad (5.28)$$

From the last expression we obtain the position of the wall:

$$\xi_w = \int_0^{-\phi_w} \frac{dz}{\sqrt{q^2 + 2p\left(\sqrt{1+2z} + e^{-z} - 2\right)}}. \quad (5.29)$$

Other dimensionless quantities are

$$\bar{n}_e = e^\phi, \quad (5.30)$$

$$\bar{n}_i = \frac{1}{\sqrt{1-2\phi}}, \quad (5.31)$$

$$\bar{v}_i = \sqrt{1-2\phi}. \quad (5.32)$$

Our calculations for the numerical parameters chosen above (for ions $Ar_2^+$) give the following results:

for Eq. (5.24): $p = 1609$;

for Eq. (5.27): $\phi_w = -5.026$;

for Eq. (5.29): $\xi_w = 0.344$;

for Eq. (5.18): $h^* = 0.017$ m;

the potential drop in the sheath: $\varphi_w = -5.026$ V;

the wall coordinate: $x_w = 0.344\, h^* = 5.72 \cdot 10^{-3}$ m;

the Hopf shift: $h = (1 - 0.344) h^* = 0.011$ m.

Fig. 9, 10 illustrate the behavior of the electric potential, as well as the density of ions and electrons in the sheath (quasineutral plasma is located on the left in the figures). Common tangent to the density graphs at the point of plasma quasi-neutrality in Fig. 10 shows the position of the Hopf point H(1, 0) (in the lower right corner of the image), which determines the Hopf shift $h$ to determine the Robin boundary condition for the equation of ambipolar diffusion in plasma.

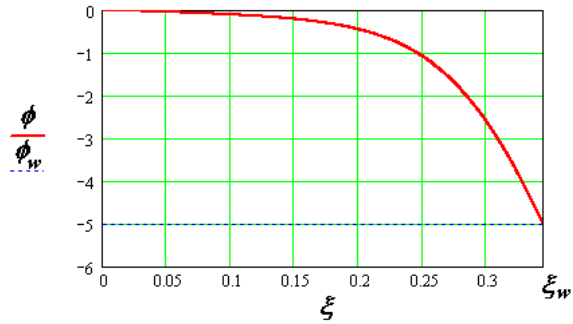

**Fig. 9.** Electric potential, volt, in the sheath. $h^* = 0.017$ m is selected as a unit of length.

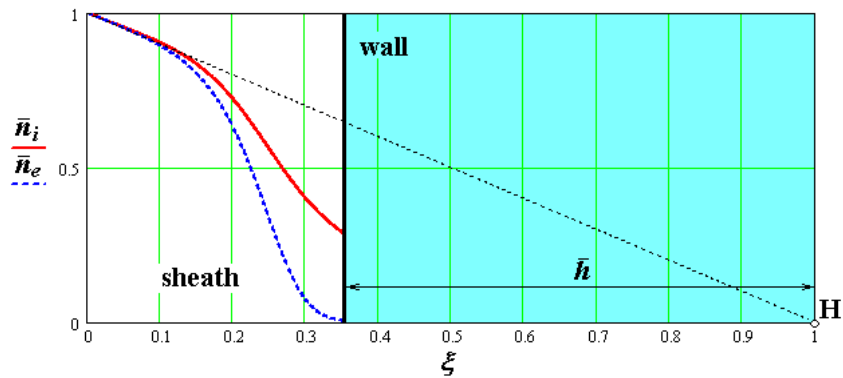

**Fig. 10.** Ion and electron densities (dimensionless). $\bar{h} = h/h^*$.

This example illustrates the possibility of using rather complex information to determine BC in the absence of a solution to the kinetic problem for the distribution functions of ions and electrons near the wall, which were evaluated here as "Maxwell-Boltzmann electrons" and "cold ions".

## V. FEATURES OF USING THE LTTA IN OBTAINING BOUNDARY CONDITIONS

In view of the generally accepted use of LTTA in the description of electrons [4], [28], [29], we return to it for further discussion.

Recall that the Maxwell LTTA consists in the approximation

$$f(\mathbf{r},\mathbf{v}) = F_T(v)\left(n(\mathbf{r}) + \frac{m}{kT}\mathbf{\Gamma}(\mathbf{r})\cdot\mathbf{v}\right), \quad F_T(v) = \left(\frac{m}{2\pi kT}\right)^{\frac{3}{2}}\exp\left(-\frac{mv^2}{2kT}\right), \quad (26)$$

where $n(\mathbf{r})$ is the particle density, and

$$\mathbf{\Gamma}(\mathbf{r}) = -D\nabla n(\mathbf{r}) + \mathrm{sgn}(q)\mu n \mathbf{E}(\mathbf{r}) \quad (27)$$

is the vector of the related flux density. Non-Maxwell LTTA, which is often used to describe thermally non-equilibrium discharges for electrons, is different from the Eq. (26):

$$f(\mathbf{r},\mathbf{v}) = n(\mathbf{r})F(v) - \frac{1}{\omega}\left(F(v)\nabla n(\mathbf{r}) + \frac{1}{v}F'(v)n(\mathbf{r})\frac{q}{m}\mathbf{E}\right)\cdot\mathbf{v}, \quad (28)$$

$$\mathbf{\Gamma} = -\frac{1}{3}\left(\left\langle\frac{v^2}{\omega}\right\rangle\nabla n + \left\langle\frac{F'}{F}\frac{v}{\omega}\right\rangle\frac{q}{m}n\mathbf{E}\right). \quad (29)$$

Here, $\omega$, s$^{-1}$, is the relaxation rate of the non-uniform distribution $f$ to some steady distribution $F$. The angle brackets denote the distribution average of $F(v)$, and the "prime" is a derivative. The function $F$ of one argument, velocity or kinetic energy, is determined from the one-dimensional energy balance equation [10].

As we showed above, the use of LTTA in deriving the boundary conditions for the density of some plasma component is not necessary. The main premise is to separate the particle stream into two types: those going *to* the wall and those going *from* the wall. Moreover, everyone understands[14] that for the correct mathematical use of LTTA, the first and second terms in the LTTA must have different scales; namely, the value of the second, anisotropic, term must be *much less* than the value of the first, isotropic, term. But the two-stream approach leads to the alignment of these scales on the absorbing wall. Thus, a good two-stream idea (split the stream in two) does not look very good, namely when using LTTA. Avoiding the use of LTTA, we maintain a sound foundation and discard all unnecessary.

## VI. SUMMARY (CONCLUSIONS)

We have demonstrated how the two-stream idea can be properly generalized to provide a general description of BC for all kinds of particles. In this generalization, we used the method of consequent derivation, and we deviate from Hagelaar and others [18] in that we avoid intuitive statements. We have shown that their formulas[15], including the "jumping" quantity $a_e$, are incorrect. It is worth noting that the popular mathematical package COMSOL 5.0-5.4 still uses these erroneous formulas for boundary conditions in plasma applications[16].

---

[14] See [30] page 79, for example.

The goal of this study is to clarify the formulation of the boundary conditions for the equations of plasma hydrodynamics on the (partially) absorbing and (possibly) emitting walls. These conditions are the Robin (or 3$^{rd}$ kind) conditions for particle density. They are uniform in the absence of wall emission, and non-uniform in the presence of wall emission. In particular, the Dirichlet or Neumann condition can be obtained if the Hopf shift tends to zero or infinity, respectively.

We have shown that in the general case (without using LTTA), the average of the absolute value of the normal component of the particle velocity on the wall $\overline{|v_x|}$ (see Eq. (6)) is required to formulate the boundary condition. Therefore, our boundary conditions will be characterized by the quality that we can provide this quantity.

We introduce the concepts of *Hopf velocity*, *Hopf shift*, *Hopf emission density* and the *Hopf point*, inheriting the concept of the *Hopf constant* $c_{\text{Hopf}} = h/\lambda_{\text{free}}$, which has arisen earlier as a solution to the Milne problem [32]. In many aspects, the Milne problem is similar to the plasma boundary problem of the highest level of understanding. These new concepts are at the same time brief in wording, and give us the opportunity to express our sincere gratitude to the researcher [33], who was the first to formulate and solve the kinetic (Boltzmann) equation for an absorbing (or radiation-transparent) wall. For plasma walls a solution to this problem is forthcoming.

ACKNOWLEDGMENTS

This research has been financially supported by the National Natural Science Foundation of China under Grant No. 11775062.

APPENDIX: ABOUT THE REFLECTION COEFFICIENT

The two-stream approach developed in Chapter II is implemented most consistently at the kinetic level. Let us separate the distribution function of some sort of particles near the wall into two terms[17]:

---

[15] See [18] below Eq. (11).
[16] See [31] for details.
[17] The value of the distribution on the set $v_x = 0$ does not matter, because it is a set of zero measure.

$$f(\mathbf{v}) = f_+(\mathbf{v}) + f_-(\mathbf{v}),$$

$$f_+(\mathbf{v}) = \begin{cases} f(\mathbf{v}), & v_x > 0; \\ 0, & v_x < 0. \end{cases} \tag{1A}$$

$$f_-(\mathbf{v}) = \begin{cases} f(\mathbf{v}), & v_x < 0; \\ 0, & v_x > 0. \end{cases}$$

Using this separation, we enable to replace the expression for reflection in Eq. (5) with a more detailed and more reasonable expression

$$f_-(\mathbf{v}) = \int_{v'_x > 0} d^3v' B(\mathbf{v}, \mathbf{v}') v'_x f_+(\mathbf{v}') + f_{em}(\mathbf{v}). \tag{2A}$$

$$B(\mathbf{v}, \mathbf{v}')\big|_{v_x > 0} = 0, \quad f_{em}(\mathbf{v})\big|_{v_x > 0} = 0. \tag{3A}$$

Eq. (2A) is the most general form of the two-stream approach. It includes the possibility of any way of scattering a particle on a wall surface: elastic, inelastic, in a random direction, etc. An essential feature is that the kernel $B$ of the linear reflection operator — the *scattering indicatrix* — depends on the properties of the interaction of the particle with the surface and does not depend on the distribution function of particles near the wall. From Eq. (2A) we can obtain

$$\Gamma_- = \int_{v_x < 0} d^3 v f_-(\mathbf{v})(-v_x) = \int_{v_x < 0} d^3 v (-v_x) \int_{v'_x > 0} d^3 v' B(\mathbf{v}, \mathbf{v}') v'_x f_+(\mathbf{v}') + \int_{v_x < 0} d^3 v (-v_x) f_{em}(\mathbf{v}). \tag{4A}$$

$$\Gamma_- = \int_{v_x > 0} d^3 v B_1(\mathbf{v}) v_x f_+(\mathbf{v}) + \Gamma_{em}, \tag{5A}$$

$$B_1(\mathbf{v}) = \int_{v'_x < 0} d^3 v' (-v'_x) B(\mathbf{v}', \mathbf{v}). \tag{6A}$$

Comparing Eq. (5) and Eq. (5A), we can obtain

$$R = \frac{\int_{v_x > 0} d^3 v B_1(\mathbf{v}) v_x f_+(\mathbf{v})}{\int_{v_x > 0} d^3 v\, v_x f_+(\mathbf{v})}. \tag{7A}$$

As you can see, in the general case, the reflection coefficient $R$ also depends on the distribution function of the particles at the wall. An important exception is the situation when the integral (6A) is independent of the velocity vector. Then $R = B_1 = \text{const}$. What are examples of such a situation? –

(a) Absolutely absorbing wall: $B(\mathbf{v}, \mathbf{v}') \equiv 0$. Then $R = B_1 = 0$.

(b) *Mirror elastic* reflection of particles:

$$B(\mathbf{v}, \mathbf{v}')(\mathbf{n} \cdot \mathbf{v}') = \delta^3\left(\mathbf{v} - (\mathbf{v}' - 2\mathbf{n}(\mathbf{n} \cdot \mathbf{v}'))\right), \quad (\mathbf{n} \cdot \mathbf{v}') \equiv v'_x.$$

Then $B_1(\mathbf{v}) = \int_{v'_x < 0} d^3 v' \frac{-(\mathbf{n} \cdot \mathbf{v}')}{(\mathbf{n} \cdot \mathbf{v})} \delta^3\left(\mathbf{v}' - (\mathbf{v} - 2\mathbf{n}(\mathbf{n} \cdot \mathbf{v}))\right) = \theta(v_x), \quad R = 1.$

$\theta(x) = 1, x > 0; \quad \theta(x) = 0, x < 0.$

(c) *Mirror inelastic* reflection of particles, and the probability of absorption $a$:

$$B(\mathbf{v},\mathbf{v}')(\mathbf{n}\cdot\mathbf{v}') = \frac{1-a}{c}\delta^3\big(\mathbf{v}-(\mathbf{v}'-(1+c)\mathbf{n}(\mathbf{n}\cdot\mathbf{v}'))\big), \quad (\mathbf{n}\cdot\mathbf{v}') \equiv v'_x, \quad 0 < c < 1.$$

$$B_1(\mathbf{v}) = \int_{v'_x<0} d^3v' \frac{-(\mathbf{n}\cdot\mathbf{v}')}{(\mathbf{n}\cdot\mathbf{v})}\frac{1-a}{c}\delta^3\big(\mathbf{v}'-(\mathbf{v}-(1+c)\mathbf{n}(\mathbf{n}\cdot\mathbf{v}))\big) = \theta(v_x)(1-a), \quad R = 1-a.$$

(d) *Diffuse elastic* reflection of particles:

$$B(\mathbf{v},\mathbf{v}') = \frac{1}{\pi v^3}\delta(v-v'),$$

$$B_1(\mathbf{v}) = -\int_{v'_x<0} d^3v' \frac{(\mathbf{n}\cdot\mathbf{v}')}{\pi v^3}\delta(v-v') = 1.$$

(e) *Diffuse inelastic* reflection of particles, and the probability of absorption $a$:

$$B(\mathbf{v},\mathbf{v}') = \frac{1-a}{\pi v^3}\delta(v-cv'), \quad 0 < c < 1.$$

$$B_1(\mathbf{v}) = -(1-a)\int_{v'_x<0} d^3v' \frac{(\mathbf{n}\cdot\mathbf{v}')}{\pi v'^3}\delta(cv-v') = 1-a. \quad R = 1-a.$$

All these examples assume *constant* parameters $c$ and $a$. The dependence of the scattering and absorption parameters on the velocity can cause the dependence of the reflection coefficient $R$ on the particle distribution function near the wall.

Now let's look at a special question: how the results of Chapter II can be used for the so-called "electron energy fluid" for an extended model of three-fluid plasma, which includes positive ions, electrons, and electron energy as three liquid components that obey the equations of drift and diffusion. Formally, nothing prevents the derivation of all the formulas in this chapter using the *energy-fluid distribution function* (EFDF):

$$f_\varepsilon(\mathbf{r},\mathbf{v}) \equiv \frac{m_e v^2}{2} f_e(\mathbf{r},\mathbf{v}). \tag{8A}$$

Multiplying Eq. (28) by $\frac{m_e v^2}{2}\mathbf{v}$, and integrating over the space of velocities, we can obtain

$$\boldsymbol{\Gamma}_\varepsilon(\mathbf{r}) = \int d^3v f_\varepsilon(\mathbf{r},\mathbf{v})\mathbf{v} = n(\mathbf{r})\int d^3v \frac{m_e v^2}{2}\mathbf{v}F(v) - \int d^3v \frac{1}{\omega}\frac{m_e v^2}{2}\mathbf{v}\left(F(v)\nabla n(\mathbf{r}) + \frac{1}{v}F'(v)n(\mathbf{r})\frac{q}{m}\mathbf{E}\right)\cdot\mathbf{v},$$

$$\boldsymbol{\Gamma}_\varepsilon(\mathbf{r}) = \int d^3v f_\varepsilon(\mathbf{r},\mathbf{v})\mathbf{v} = -\frac{1}{3}\int_0^\infty 4\pi v^2 dv \frac{1}{\omega}\frac{mv^2}{2}v^2\left(F(v)\nabla n(\mathbf{r}) + \frac{1}{v}F'(v)n(\mathbf{r})\frac{q}{m}\mathbf{E}\right),$$

$$\Gamma_\varepsilon(\mathbf{r}) = \int d^3 v f_\varepsilon(\mathbf{r},\mathbf{v})\mathbf{v} = -D_\varepsilon \nabla n(\mathbf{r}) + \mu_\varepsilon n(\mathbf{r})\mathbf{E},$$

$$D_\varepsilon = \frac{2\pi}{3}\int_0^\infty dv \frac{m_e v^6}{\omega} F(v) = \frac{m_e}{6}\left\langle \frac{v^4}{\omega}\right\rangle = \frac{1}{3}\left\langle \frac{v^2}{\omega}\right\rangle_\varepsilon, \qquad (9A)$$

$$\mu_\varepsilon = \frac{2\pi}{3}\int_0^\infty dv \frac{ev^5}{\omega} F'(v) = \frac{e}{6}\left\langle \frac{v^3}{\omega}\frac{F'(v)}{F(v)}\right\rangle = \frac{1}{3}\frac{e}{m_e}\left\langle \frac{v}{\omega}\frac{F'(v)}{F(v)}\right\rangle_\varepsilon.$$

As you can see, these expressions show good agreement with Eq. (29):

$$\Gamma = -\frac{1}{3}\left(\left\langle \frac{v^2}{\omega}\right\rangle \nabla n + \left\langle \frac{F'}{F}\frac{v}{\omega}\right\rangle \frac{q}{m} n\mathbf{E}\right).$$

That is, the drift-diffusion model for the "energy electron fluid" is available. But what about Eq. (5)? – To understand this, let's use Eq. (4A):

$$\Gamma_-^{(\varepsilon)} = \int_{v_x<0} d^3 v f_\varepsilon(\mathbf{v})(-v_x) = \int_{v_x<0} d^3 v f_e(\mathbf{v})\frac{m_e v^2}{2}(-v_x) =$$

$$= \int_{v_x<0} d^3 v(-v_x)\int_{v'_x>0} d^3 v' \frac{v^2}{v'^2} B(\mathbf{v},\mathbf{v}')v'_x \frac{m_e v'^2}{2}f_e(\mathbf{v}') + \int_{v_x<0} d^3 v(-v_x)\frac{m_e v^2}{2}f_{em}(\mathbf{v}).$$

$$\Gamma_-^{(\varepsilon)} = \int_{v_x<0} d^3 v(-v_x)\int_{v'_x>0} d^3 v' \frac{v^2}{v'^2} B(\mathbf{v},\mathbf{v}')v'_x f_\varepsilon(\mathbf{v}') + \Gamma_{em}^{(\varepsilon)},$$

$$\Gamma_-^{(\varepsilon)} = \int_{v_x>0} d^3 v B_1^{(\varepsilon)}(\mathbf{v})v_x f_\varepsilon(\mathbf{v}) + \Gamma_{em}^{(\varepsilon)}, \qquad (10A)$$

$$B_1^{(\varepsilon)}(\mathbf{v}) \equiv \int_{v'_x<0} d^3 v'(-v'_x)\frac{v'^2}{v^2} B(\mathbf{v}',\mathbf{v}), \qquad (11A)$$

$$\Gamma_{em}^{(\varepsilon)} \equiv \int_{v_x<0} d^3 v(-v_x)\frac{m_e v^2}{2} f_{em}(\mathbf{v}). \qquad (12A)$$

$$R_\varepsilon = \frac{\int_{v_x>0} d^3 v B_1^{(\varepsilon)}(\mathbf{v})v_x f_\varepsilon(\mathbf{v})}{\int_{v_x>0} d^3 v\, v_x f_\varepsilon(\mathbf{v})}. \qquad (13A)$$

As can be seen from the expressions obtained, the energy reflection coefficient can be independent of the electron distribution function in all cases that correspond to the above examples with constant parameters. The "absorption" of the "energy electron fluid" by the wall can cause some difference in expressions (6A) and (11A) in situation where all electrons are reflected, but the reflection is inelastic: $R=1$, $R_\varepsilon = c^2 < 1$. This is quite natural for inelastic scattering: the number of particles is conserved, but the energy is not. In other aspects, the derivation of formulas of Chapter II for electron energy fluid is quite similar.

_______________

1. https://www.encyclopediaofmath.org/index.php/Hydrodynamic_approximation
2. G.N. Hays, C.J. Tracy, H.J. Oskam, J. Chem. Phys **60**, 2027 (1974).
3. A. Wilson, B. Shotorban, Phys. of Plasmas **25**, 053509 (2018).
4. J.P. Boeuf, L.C. Pitchford, Phys. Rev. E **51**, 1376 (1995).
5. V.V. Gorin, European Phys. J. D **59**, 241 (2010), DOI: 10.1140.
6. L.D. Landau, E.M. Lifshitz, *Theoretical physics* (Pergamon Press, Oxford, New York, 1984) Vol. 6.
7. A. Derzsi, P. Hartmann, I. Korolov, J. Karacsony, G. Bano, Z. Donko, J. Phys. D: Appl. Phys. **42**, 225204 (2009).
8. I.P. Shkarofsky, T.W. Johnston, M.P. Bachynski, *The particle kinetics of plasmas* (Addison-Wesley, London, 1966).
9. Yu.P. Raizer, *Gas Discharge Physics* (Hardcover 2001).
10. G.J.M. Hagelaar, L.C. Pitchford, Plasma Sources Sci. Technol. **14**, 722 (2005).
11. A.A. Kudryavtsev, A.V. Morin, L.D. Tsendin, Technical Phys. **78**, 8, 77 (2008).
12. U. Kortshagen, C. Busch, L.D. Tsendin, Plasma Sources Sci. Technol. **5**, 1 (1996).
13. https://www.encyclopediaofmath.org/index.php/Third_boundary_value_problem
14. E. A. Milne, Mon. Notices Roy. Astron. Soc. **81**, 361 (1921).
15. B. Davison, *Neutron transport theory* (Clarendon, Oxford, England, 1957).
16. V.V. Gorin, WSEAS Transactions on Heat and Mass Transfer **12**, 144 (2017).
17. E.W. McDaniel, *Collision phenomena in ionized gases* (Willey, New York, 1964).
18. G.J.M. Hagelaar, F.J. deHoog, G.M.W. Kroesen, Phys. Rev. E **62**, 1, 1452 (2000).
19. S.I. Eliseev, A.A. Kudryavtsev, H. Liu, Z. Ning, D. Yu, A.S. Chirtsov, IEEE Transact. on Plasma Sci. **44**, 11, 2536 (2016).
20. I. Rafatov, E.A. Bogdanov, A.A. Kudryavtsev, Phys. Plasmas **19**, 033502 (2012).
21. A.V. Phelps, J. Res. Natl. Inst. Stand. Technol. **95**, 407 (1990).
22. P.J. Chantry, A.V. Phelps, G.J. Schulz, Phys. Rev. **152**, 12, 81 (1966).
23. https://www.encyclopediaofmath.org/index.php/Boundary_value_problem,_ordinary_differential_equations
24. https://www.encyclopediaofmath.org/index.php/Differential_equations_with_small_parameter
25. https://www.encyclopediaofmath.org/index.php/Cramer_rule
26. https://www.encyclopediaofmath.org/index.php/Boltzmann_distribution
27. K.-U. Riemann, J. Phys. D: Appl. Phys. **24**, 493 (1991).
28. L.L. Alves, Plasma Sources Sci. Technol. **16**, 557 (2007).

29. L.L. Alves, G. Gousset, C.M. Ferreira, Phys. Rev. E **55**, 1, 890 (1997).
30. A.V. Rozhansky, L.D. Tsendin, *Transport Phenomena in Partially Ionized Plasma* (Taylor & Francis, London, 2001).
31. COMSOL Multiphysics Reference Manual, version 5.4, COMSOL, Inc.
32. https://www.encyclopediaofmath.org/index.php/Milne_problem
33. E. Hopf, *Mathematical problems of radiative equilibrium* (Camb., 1934).